\documentclass[doublecol]{epl2} 
\usepackage{soul}
\usepackage{color}
\usepackage{graphicx}  
\usepackage{dcolumn}   
\usepackage{bm}        
\usepackage{amssymb}   
\usepackage{epstopdf}
\usepackage[normalem]{ulem}

\title{Scale-free networks may not necessarily witness cooperation}

\author{Deep Nath\inst{1}, Saptarshi Sinha\inst{1} \and Soumen roy\inst{1}\footnote{soumen@jcbose.ac.in}}
\shortauthor{D. Nath\etal}
\institute{\inst{1}{Department of Physics, Bose Institute, 93/1 Acharya Prafulla Chandra Road, Kolkata 700009, India}}

\abstract{Networks with a scale-free degree distribution are widely thought to promote cooperation in various games. Herein, by studying the well-known prisoner's dilemma game, we demonstrate that this need not necessarily be true. For the very same degree sequence and degree distribution, we present a variety of possible behaviour. We reassess the perceived importance of  hubs in a network towards the maintenance of cooperation. We also reevaluate the dependence of cooperation on network clustering and assortativity.}

\begin{document}

\maketitle

Evolutionary game theory (EGT) has captured the serious attention of evolutionary biologists, ecologists, computer scientists and statistical physicists over the last few decades. This is chiefly due to its potential to effectively understand the challenge of evolution and maintenance of cooperation from microscopic to macroscopic {scales} in the Darwinian context.

In classical game theory, players are rational individuals who can choose their own strategy \cite{morgenstern1944theory}. This decision-making ability enables players to maximize their payoff in games. However, EGT differs from classical game theory in that individuals are not driven by rationality {\em per se} and may not require complete information about other players \cite{Dong2020}. Here, players are genetically constrained to perform a specific strategy \cite{Taylor1978}, which naturally discourages ``mixing" of strategies. Such invariant strategies are therefore also referred to as pure strategies. EGT  deals with interactions between two or more genetically distinct populations sharing common resources and other environmental factors. One of the goals of EGT is to model population dynamics on evolutionary time scales. Here, the population structure in the steady-state is comprised of players having evolutionary stable strategies (ESS). Evolutionary stability refers to such genetic compositions in which no other mutant genotype can successfully invade a population by evolutionary processes like natural selection \cite{Nowak2006,Szabo2007,Sinha2019}.

EGT enables us to investigate various evolutionary processes by knowing the frequency-dependent steady-state outcome of two or more interacting populations. Prime factors in evolutionary games include the strategy of players and game rules.  Cooperation between living organisms may flourish irrespective of the presence of free-riders \cite{cheney2011extent,sachs2012origins,mckenna2015simultaneous}. EGT has been studied primarily on four types of games: prisoner's dilemma (PD), harmony, snowdrift and coordination, which differ in their payoff values and steady states \cite{Szabo2007}. In the last few decades, much research has been done on the maintenance of cooperation in various games. Among these, PD is significant because defection would be the natural tendency in PD\cite{Perc2010}.%

Apart from game rules, the underlying structure of the population also plays an essential role in the outcome of the game \cite{nowak1992evolutionary,szolnoki2010reward,antonioni2013coordination, lee2011emergent, epidemic, maciejewski2014evolutionary, quorum, Gomez2007}. The underlying graph topology imparts spatial restrictions on the interactions between players. These spatial restrictions may act in favour of cooperation. In PD games played on homogeneous population structures, it is difficult to maintain cooperation  \cite{Sinha2019,szolnoki2008towards}. On the other hand, cooperation could thrive in heterogeneous populations. Thus, the outcome of a game depends on the structure of the population, types of payoffs and sundry factors like mobility \cite{wu2009diversity}. 

Networks  have been found to be useful in fields \cite{ba2002rmp, newman2003structure} as diverse as mutagenesis and phage resistance \cite{sinha2020PRM}, image-processing and non-invasive diagnostics \cite{banerjee2015using}, infrastructure \cite{pre151} and optogenetics\cite{kaur2015mapping,deb2020residue}.  While degree is only one of the many metrics in networks \cite{ba2002rmp, newman2003structure}, it has received perhaps the most emphasis in network literature \cite{roy2012systems}.  Graphs with power-law degree distributions have been generally alluded to as ``scale-free networks" in {the} literature  \cite{ba1999sf, roy2012systems, broido2019scale, clauset2009power}. Heterogeneity in scale-free (SF) networks can be better understood through measures such as the S-metric \cite{Doyle2005, Tsiotas6701}. As {is} well-known, the mechanism of generation \cite{ba1999sf,d2007emergence} can imprint its signature on the structure of the network \cite{Doyle2005}.

It has been reported earlier that scale-free networks possess an inherent tendency to promote cooperation \cite{santos2005scale}.  The underlying intuition seems to be that when cooperators are hubs, they can survive in a population by accumulating higher payoffs as compared to their  defecting neighbours \cite{Santos2006,Santos2008}. It has also been thought that factors like clustering {and} assortativity could influence this outcome, as {a} higher clustering coefficient and high assortativity between cooperators may enhance cooperation \cite{Assenza2008,Kuperman2012,Wang2010,Wang2012,Smith2018}. Herein, we demonstrate that these need not necessarily be true.  Indeed, for the very same degree sequence and degree distribution, we demonstrate that SF networks may display a rich diversity in behaviour with regard to cooperation. 

Let $\cal {G(V,E)}$ denote a graph,  where $\cal V$ and $\cal E$ denote the set of nodes and edges respectively.  $|{\cal V}|=\cal N$  and $|\cal E|$ denotes the number of nodes and edges respectively in $\cal {G(V,E)}$. Henceforth, we often refer to $\cal {G(V,E)}$ as $\cal {G}$. We now define
\begin{equation}
S=\frac{s({\cal G})}{s_{max}}=\frac{\sum_{{\cal E}_{ij} \in {\cal E}}{k_i}{k_j}}{s_{max}}
\label{eq:s}
\end{equation}
Here, $i$ and $j$ are the end nodes of the edge ${\cal E}_{ij} \in {\cal E}$.  The degree of node $i$ and $j$ is denoted by $k_i$ and $k_j$ respectively. 
If $\cal K$ denotes the degree sequence of ${\cal G}$, let ${\cal G} (\cal K)$ denote the {\em set of graphs} with degree sequence $\cal K$. $s_{max} = max \{s({\cal G}) : {\cal G} \in {\cal G}(\cal K)\}$, whence  $0 < S({\cal G}) \le 1$. Only a completely disconnected graph has $S=0$ and is therefore excluded herein. If graphs having different values of $S({\cal G})$ possess the same degree sequence -- their degree distribution is obviously  identical. Herein, $S({\cal G})$ is used to represent different graphs with identical degree sequence and hence identical degree distribution \cite{Doyle2005}. Henceforth, we mostly refer to $S({\cal G})$  simply as $S$. $S({\cal G})$ can be defined for virtually any graph. However, its usefulness  is readily apparent and has been widely used to differentiate between various SF networks  \cite{Doyle2005,Li2005}, which is the prime object of study in this letter.

Degree assortativity, $r$, broadly captures whether  nodes having similar degree are connected to each other \cite{Li2005, Newman2002}. 
\begin{equation}
r=\frac{[\sum_{{\cal E}_{ij} \in {\cal E}}{k_i}{k_j}]-[{\sum_{i \in {\cal V}} \frac{{k_i}^2}{2}}]^2 /|{\cal E}|}{[{\sum_{i \in {\cal V}} \frac{{k_i}^3}{2}}]-[{\sum_{i \in {\cal V}} \frac{{k_i}^2}{2}}]^2 /|{\cal E}|}
\label{eq:r}
\end{equation}

As well known,  $-1 \le r \le 1$.  Graphs with positive and negative values of $r$ are termed assortative and disassortative respectively. In assortative graphs, nodes with higher degree are predominantly connected to each other. In disassortative graphs, nodes with higher degree are predominantly connected to nodes with lower degree. 

$S (\cal G)$ reflects the extent to which a given graph is scale-free \cite{Doyle2005, Li2005}. $\forall\cal G \in \{\cal G(\cal K)\}$ possess a strictly identical degree sequence, by definition.  Herein, we are {\em not only} interested in graphs with the same power-law degree distribution, but we are additionally interested in graphs with an {\em identical degree sequence}. Therefore,  in this letter $S$ and not $r$ is the natural and obvious choice to perform the role of the key structural index.

We simulate the evolutionary PD game on heterogeneous populations. The population structure has been initially considered as a Barab{\'a}si-Albert (BA) network. {BA networks can be generated through  the mechanism of preferential attachment \cite{ba1999sf,d2007emergence}. They exhibit a power-law degree distribution.} The extent of prevalence of scale-free networks in the real world has been extensively discussed \cite{broido2019scale, holme2019rare}. For each ensemble, initially a BA network, $\cal G_{BA}$, is generated. From $\cal G_{BA}$, a {\em set of scale-free networks}, $\{\cal G_{SF}\}$ is obtained by repeated {\em degree preserving double-edge swap} \cite{Maslov910, PhysRevE.70.066102}. Thus, at every step, the removal of two randomly chosen edges, ${\cal E}_{ij}$ and ${\cal E}_{kl}$,  is accompanied by the creation of two new edges, ${\cal E}_{ik}$ and ${\cal E}_{jl}$, while retaining the degree of each node. We can hardly overemphasise that $\forall \cal G\in\{\cal G_{SF}\}$ have the same degree sequence and naturally their  degree distribution is identical to that of $\cal G_{BA}$. $\forall \cal G\in\{\cal G_{SF}\}$ would obviously possess a value of $S$ different from $S(\cal G_{BA})$.  It should be noted that no node or edge is removed or added during rewiring by degree preserving double-edge swaps. 

We can easily obtain the value of $max(s)$ in $\{\cal G_{SF}\}$. However it is not possible to achieve an arbitrarily specified low value of $S,  \forall \cal G\in\{\cal G_{SF}\}$. The minimum obtainable value of $S$ would depend on $\cal N$, $\cal E$, and edge density of $\cal G_{BA}$ among other factors. Here, ${\cal N}=1024$ and we have been able to generate graphs with $S$ as low as $S=0.3$. Besides, generating graphs with arbitrarily low values of $S$ at a given $\cal N$ is computationally inhibitive \cite{Doyle2005}.  

At the start of each ensemble, the population is randomly  divided into an equal number of cooperators, $C$, and defectors, $D$. Thus, the {\em initial fraction} of cooperators, $f_{C_i}= 0.5$. Each node in ${\cal G}$ represents a player, who can interact with other players directly connected to it. Here, the strategies of the players and  rules of the game do not affect the population structure, irrespective of whether the underlying network is SF or not \cite{broido2019scale, holme2019rare}.  Of course, recently it has been thought that the emergence of SF networks can depend upon the proportions of different types of players present. {Indeed, in some agent-based modeling frameworks, agents influence the fundamental nature of the network upon which they act, including emergence of scale-free behavior, even for a fixed set of interaction rules \cite{fleming2021scale}.}%

Interaction between two cooperators results in a reward, $\cal R$. If two defectors interact with each other, they will earn punishment, $\cal P$. On the other hand, interaction between $C$ and $D$ will lead to sucker's payoff, $\cal S$, for $C$ and temptation, $\cal T$, for $D$. In a PD game, ${\cal T}>{\cal R}>{\cal P}>{\cal S}$ \cite{Szabo2007}. Herein, these payoff values are considered to be ${\cal R}=1.0$, $1.0<\cal{T} \leq$ $2.0$, ${\cal P}=0.0$ and ${\cal S}=0.0$ \cite{santos2005scale}. In each round, both transient and counting time incorporates payoff determination and strategy upgradation processes. Initially, players would accumulate payoff depending on interaction with their neighbors. If an individual, $i$, interacts with a randomly chosen neighbor, its payoff is $\pi_{ij}$. Generally $\pi_{ij} \ne \pi_{ji}$. The value of $\pi_{ij}$ would be $\cal R$, $\cal T$, $\cal P$ or $\cal S$. The accumulated payoff of $i$ is $\Pi_i= \sum_{j} \pi_{ij}$. After payoff determination, individuals will update their strategy synchronously.
Let $\Pi_i$ and $\Pi_j$ denote the accumulated payoffs of $i$ and $j$ respectively. $i$ will imitate the strategy of $j$ with a probability,
\begin{equation}
{P}_{i\rightarrow j}=\frac{\Pi_j-\Pi_i}{({\cal T}-{\cal S}) \times max(k_i, k_j)} \Theta_{\Pi_j \textgreater \Pi_i}
\label{eq:P}
\end{equation}
Here $\Theta_{\Pi_j \textgreater \Pi_i}=1$ for $\Pi_j \textgreater \Pi_i$ and zero otherwise. This condition indicates that individuals will try to maximize their payoff and $i$ will imitate $j$'s strategy only if $\Pi_j \textgreater \Pi_i$.
$10^4$ generations of transient time have been considered in each ensemble. The final fraction of cooperators, $f_C$, is averaged over $10^3$ generations. For each network, ${\cal N}=1024$ and average degree, $\langle k \rangle =4$. 

\begin{figure}[htbp]
\centering
\includegraphics[width=8cm,height=5cm]{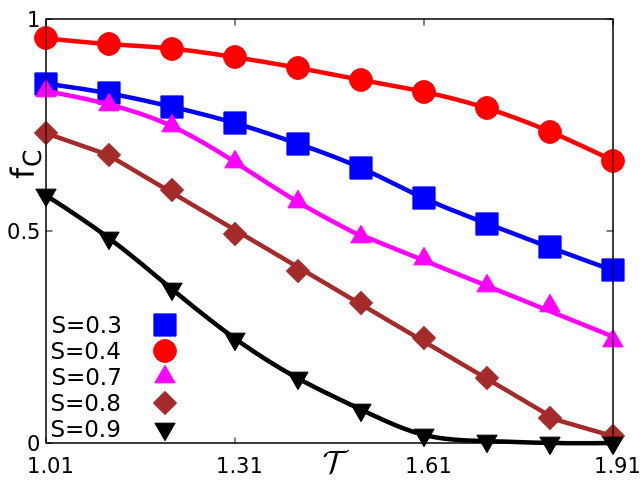}
\caption{Fraction of cooperators, $f_C$, versus temptation, $\cal T$, at different values of S-metric, $S$. Results are for $f_{C_i}=0.5$, ${\cal N}=1024$, $\langle k\rangle=4$, $E_{\cal N}=1600$ ensembles. Cooperation is high for $S=0.4$ at all values of $\cal T$. However, for higher and lower values of $S$ cooperation is not maintained well. The standard error is smaller than the size of the data points.} 
\label{fig:fraction}
\end{figure}

Fig. \ref{fig:fraction} presents a plot of the fraction of cooperators, $f_C$, against temptation, $\cal T$, at different values of $S$. Higher values of $\cal T$ favor defection and result in a decrease of $f_C$. We also observe that the dependence of $f_C$ on $S$ is highly non-monotonic. Further, the maintenance of cooperation is high only in and around $S=0.4$. This demonstrates that the maintenance of cooperation in scale-free networks is not decided by the degree distribution alone.

\begin{figure}[htbp]
\centering
\includegraphics[width=8cm,height=5cm]{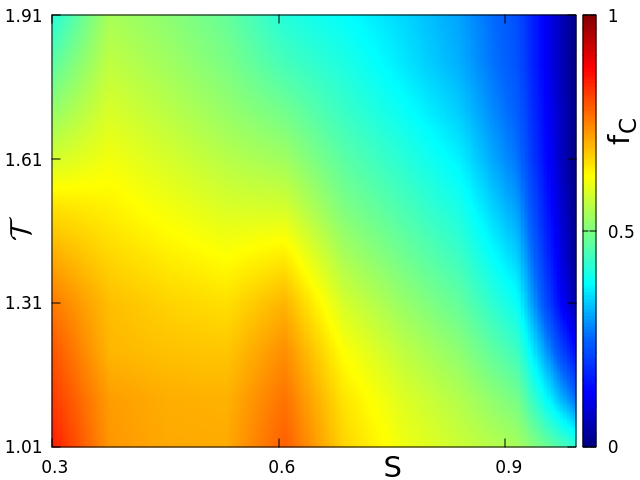}
\caption{$f_C$ versus $S$ at various values of $\cal T$ for scale-free (SF) networks. Red indicates the maintenance of cooperation and blue  its absence. Cooperation depends on both $S$ and $\cal T$. Results are for $f_{C_i}=0.5$, ${\cal N}=1024$, $\langle k\rangle=4$, $E_{\cal N}=1500$ ensembles. Cooperation is largely well-maintained or ill-maintained respectively at lower and higher values of $\cal T$ and $S$. We observe that $f_C$ is higher in and around $S=[0.35,0.45]$. This demonstrates that the maintenance of cooperation in SF networks is not decided by the degree distribution alone.} 
\label{fig:h}
\end{figure}

The complex variation of $f_C$ with $S$ at different values of $\cal{T}$ is demonstrated in Fig. \ref{fig:h}. We observe that at low values of ${\cal T}$ and $S$, cooperation is well-maintained. However, at higher values of ${\cal T}$, the maintenance of cooperation is higher in and around $S = [0.35,0.45]$.  

\begin{figure}[htbp]
\centering
\includegraphics[width=8cm,height=4.5cm]{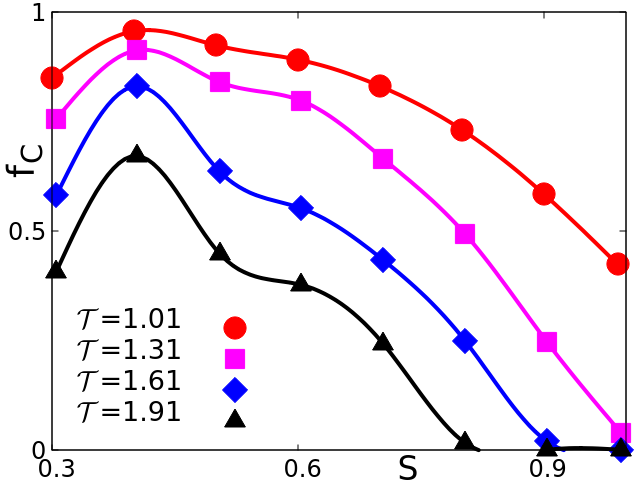}
\caption{Fraction of cooperators, $f_C$, versus $S$, at various values of $\cal{T}$.  While only scale-free graphs have been considered here -- all of them clearly do not promote cooperation. At all values of $\cal T$, $f_C$ is higher in and around $S=[0.35,0.45]$.  Here, $f_{C_i}=0.5$, ${\cal N}=1024$, $\langle k\rangle=4$, and $E_{\cal N} =1000$. The standard error is smaller than the size of the data points.} 
\label{fig:Coop}
\end{figure}

In Fig. \ref{fig:Coop}(a) we examine the behaviour of $f_C$ with respect to $S$ at various values of $\cal T$.  
We again observe that $f_C$ is higher in and around $S=[0.35,0.45]$, as witnessed earlier in Fig. \ref{fig:h}. 

It has been widely presumed that hubs are responsible for the maintenance of cooperation in heterogeneous population structures. The underlying thought seems to be that when the hubs are cooperators they can acquire higher payoffs \cite{Santos2008}. Herein, graphs having different values of $S$ possess the same degree sequence by definition. It can then be expected that $f_C$ should not depend on $S$. However, from Figs. \ref{fig:fraction}, \ref{fig:h} and \ref{fig:Coop}, it can be easily observed that $f_C$  strongly depends on $S$. 

\begin{figure}[htbp]
\centering 
\includegraphics[width=8.75cm,height=6.5cm]{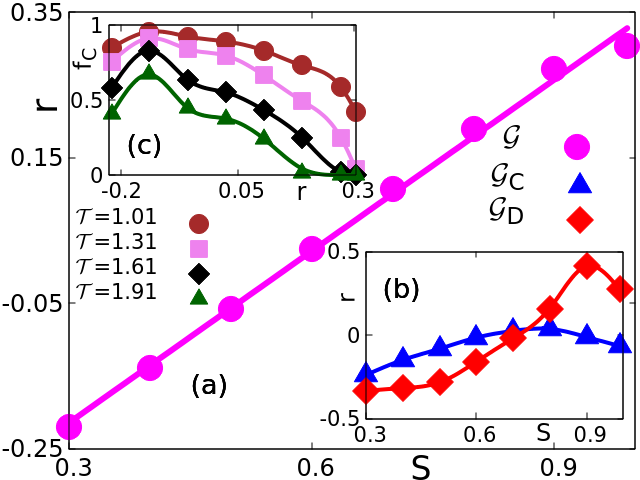}
\caption{Assortativity, $r$, versus $S$  for (a) the original graph, ${\cal G}$ and (b) cooperator graph, ${\cal G}_C$, and defector graph, ${\cal G}_D$, at ${\cal T}=1.31$. (c) Fraction of cooperators, $f_C$, versus $r$ at various values of ${\cal T}$. $f_{C_i}=0.5$, ${\cal N}=1024$, $\langle k\rangle=4$, and $E_{\cal N} =1000$. The standard error is smaller than the size of the data points.} 
\label{fig:r_all}
\end{figure}

The variation of $r$ with $S$ is studied  in Fig. \ref{fig:r_all}(a) and is observed to be consistent with reported literature \cite{Li2005}.  It has been postulated earlier {in both two-person PD games and multi-individual public goods games} that {\em assortativity among cooperators} could work in favour of cooperation \cite{Wang2012, Smith2018}. It has also been observed that if the {\em entire network is assortative} (in contrast to  assortativity among the cooperators only), it helps in maintenance of cooperation \cite{Tanimoto2010, Tanimoto2013}. {Some studies have indicated that cooperation may be sustained in disassortative networks \cite{Rong2007,Tanimoto2009}.}

When hubs act as cooperators they can accumulate higher payoffs. Hence, cooperation can be maintained in a population. {\em Also, assortativity between the hubs should operate in favor of cooperation as they can acquire higher payoffs as well.} Therefore, cooperation should be maintained in assortative graphs which possess higher values of $S$. However, in disassortative graphs cooperation might not be maintained. Since $r$ varies linearly with $S$, it would be expected that $f_C$ would possess a linear dependence on $S$. It is evident from Fig. \ref{fig:Coop} that for higher and lower values of $S$, cooperation is not maintained well enough. A suitable region for the maintenance of cooperation lies somewhere between highly assortative and highly disassortative graphs. Hence, we can conclude that networks with scale-free degree distribution do not always promote cooperation. Also, hubs and assortativity between them might not really be responsible for the maintenance of cooperation.

Assortativity among cooperators, $r_C$, can perhaps be differently scrutinised through the {\em ``cooperator graph"}, ${\cal G}_C$, instead of the original graph, ${\cal G}$ \cite{quorum}. Similarly, the {\em ``defector graph"}, ${\cal G}_D$, may be useful to understand the assortativity between defectors, $r_D$. We can construct ${\cal G}_C$ and ${\cal G}_D$ from the original graph, ${\cal G}$ \cite{quorum}. ${\cal G}_C$ and ${\cal G}_D$ are solely graphs of cooperators and defectors respectively among themselves. ${\cal G}_C$ is obtained by removing every defector and each of its connections from ${\cal G}$. Similarly ${\cal G}_D$ is obtained by pruning all cooperators and their connections from ${\cal G}$. ${\cal G}_C$ and ${\cal G}_D$ respectively capture the connectivity among cooperators and defectors themselves in ${\cal G}$,  but not between any cooperator and defector. For completeness, in Fig. \ref{fig:r_all}(b) we study the variation of $r$ versus $S$ for ${\cal G}_C$ and ${\cal G}_D$. In contrast to the linear behaviour observed in Fig. \ref{fig:r_all}(a) for the full graph, ${\cal G}$, we observe a non-linear variation in ${\cal G}_C$ and ${\cal G}_D$. We have observed earlier in Fig. \ref{fig:Coop} that the maintenance of cooperation is higher in and around $S = [0.35, 0.45]$ for ${\cal G}$. However, Fig. \ref{fig:r_all}(b) for ${\cal G}_C$ and ${\cal G}_D$ demonstrates that the value of $r_C$ is enhanced at higher values of $S$. ${\cal G}_C$  captures purely the connections between cooperators only, while, $f_C$ is calculated for the full graph, ${\cal G}$. Therefore, $f_C$ may not be really correlated with $r_C$. Also defection dominates at $S=0.99$, while $r_D$ is higher at $S=0.9$. In Fig. \ref{fig:r_all}(c), we also observe the variation of $f_C$ versus $r$ at different values of $\cal T$. In summary, the role of assortativity in the maintenance of cooperation in a population needs far larger scrutiny in order to arrive at a suitable conclusion.

\begin{figure}[htbp]
\centering
\includegraphics[width=8.75cm,height=6cm]{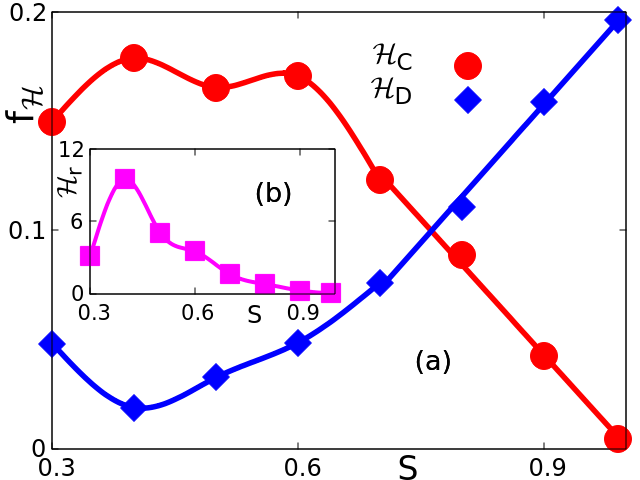}
\caption{(a) Fraction of hubs acting as cooperators, $f_{{\cal H}_C}$, or defectors, $f_{{\cal H}_D}$, and,
(b) ${\cal H}_r = f_{{\cal H}_C}/f_{{\cal H}_D}$; versus $S$ at ${\cal T}=1.31$. All graphs possess the same degree sequence and therefore the same number of hubs and an identical degree distribution. ${\cal H}_r$ peaks at $S=0.4$, alike $f_C$ in Fig. \ref{fig:Coop}. We also observe that hubs are mostly defectors at higher $S$. Results are for $\langle k\rangle=4$, $E_{\cal N}=1000$ ensembles. The standard error is smaller than the size of the data points.}
\label{fig:hub}
\end{figure}

 \begin{figure}[htbp]
\centering
\includegraphics[width=7.75cm,height=4.5cm]{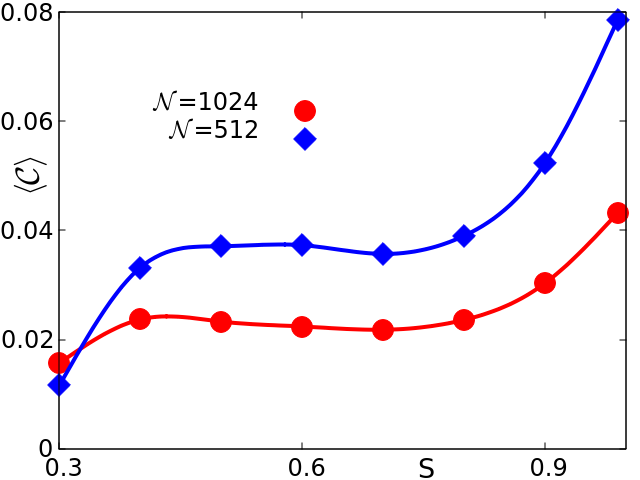}
\caption{Average clustering coefficient, $\langle {\cal C} \rangle$, versus $S$ at ${\cal T}=1.31$. Results are for (a) ${\cal N}=512$, and, (b) ${\cal N}=1024$ nodes. $\langle {\cal C} \rangle$ may not depend prominently on $S$ as $N\to\infty$. Neither has any such dependence been widely reported in literature. Figs. \ref{fig:fraction}, \ref{fig:h} and \ref{fig:Coop} exhibit a strong dependence of $f_C$ on $S$. A natural question is whether and how $f_C$ would depend on $\langle {\cal C} \rangle$, especially if the dependence of $\langle {\cal C} \rangle$ on $S$ is minimal. Results are for $\langle k\rangle=4$ and $E_{\cal N}=1200$. The standard error is smaller than the size of the data points.} 
\label{fig:av_cc_1}
\end{figure}

In Fig. \ref{fig:av_cc_1}(a), we study the average clustering coefficient, $\langle {\cal C} \rangle$, $\forall \cal G\in\{\cal G_{SF}\}$ at different values of $S$. $\langle {\cal C} \rangle$ may not prominently depend on $S$ as $N\to\infty$. Neither has any such dependence of $\langle {\cal C} \rangle$ on $S$ been widely reported in literature. We have observed earlier that Figs. \ref{fig:fraction}, \ref{fig:h} and \ref{fig:Coop} exhibit a strong dependence of $f_C$ on $S$. A natural question is whether and how  $f_C$ would depend on $\langle {\cal C} \rangle$, especially if the dependence of $\langle {\cal C} \rangle$ on $S$ is minimal. Previous studies have observed that cooperation increases with an increase in network clustering \cite{Assenza2008,Kuperman2012,Wang2010}. {Cooperation is known to decrease when ${\cal T}>2.5$, irrespective of the value of average clustering in the network \cite{Assenza2008}.}  Of course, it is also known that at higher mutation rates, even highly clustered networks may not witness cooperation \cite{Rui2010}. However, it must also be duly noted that, while the degree distribution remained unchanged in Refs. \cite{Assenza2008,Kuperman2012} -- the degree sequence likely changed. Herein, we have strictly retained the degree sequence throughout.

We now address the importance of hubs in a graph by studying the variation of the number of hubs and their clustering coefficient with $S$. It has been claimed that hubs mainly act as cooperators in a scale-free network and play an important role in maintaining cooperation \cite{Santos2006}. $\forall \cal G\in\{\cal G_{SF}\}$ possess identical degree sequence. Therefore, $\forall \cal G\in\{\cal G_{SF}\}$ can be expected to possess an identical number of hubs. Let $k_{sd}$ denote the standard deviation of the degree distribution of ${\cal G}$. Herein, we consider nodes with degree greater than $\langle k \rangle + k_{sd}$ as hubs.  Let us denote all hubs by ${\cal H}$ and those which act as cooperators and defectors by ${\cal H_C}$ and ${\cal H_D}$ respectively. The {\em number} of these hubs can then be denoted by ${\cal N_H}$, ${\cal N}_{{\cal H}_C}$ and ${\cal N}_{{\cal H}_D}$ respectively.  The respective {\em fraction} of such hubs are denoted as $f_{\cal H}$, $f_{{\cal H}_C}$ and $f_{{\cal H}_D}$. The value of $f_{\cal H}$ does not depend on the value of $S$ but is decided by $\cal K$, as aforementioned. 
In Fig. \ref{fig:hub}(a), we study the variation of $f_{{\cal H}_C}$ and $f_{{\cal H}_D}$ with $S$. We observe that as $S$ increases, $f_{{\cal H}_C}$ gradually starts declining but $f_{{\cal H}_D}$ rises.  $f_{{\cal H}_C}$ is higher at lower values of $S$ and responsible for the overall maintenance of cooperation in ${\cal G}$. However, as $S$ increases -- hubs start adopting defection. Therefore, irrespective of the presence of hubs -- cooperation is not maintained at higher values of $S$. We also study \hspace{-0.05cm}${{\cal H}_r}\hspace{-0.05cm}=\hspace{-0.05cm}{\cal N}_{{\cal H}_C}/{\cal N}_{{\cal H}_D}\hspace{-0.05cm}=\hspace{-0.05cm}f_{{\cal H}_C}/f_{{\cal H}_D}$ \hspace{-0.05cm}versus $S$ in Fig. \ref{fig:hub}(b). 
${\cal H}_r$ is highest at $S=0.4$, where cooperation is also highest as already observed in Fig. \ref{fig:Coop}. Hubs seem to play an important role in maintaining cooperation, when they are cooperators. \hspace{-0.05cm}However, whether they act as cooperators or defectors would depend on the topology of the graph.

\begin{figure}[htbp]
\centering
\includegraphics[width=8.75cm,height=7cm]{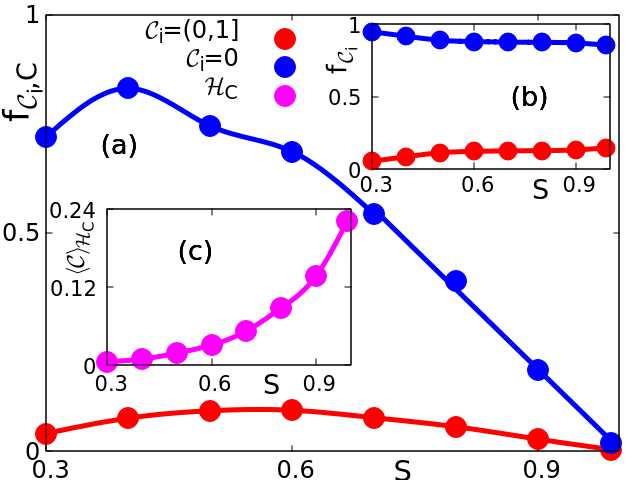}
\caption{(a) Fraction of cooperators, $f_{{\cal C}_i, C}$, possessing ${\cal C}_i=0$ and ${\cal C}_i=(0,1]$,  (b) fraction of all nodes, $f_{{\cal C}_i}$, with ${\cal C}_i=0$ and ${\cal C}_i=(0,1]$, (c) average clustering coefficient of cooperator hubs, ${\langle {\cal C} \rangle}_{{\cal H}_C}$; versus  $S$ at ${\cal T}=1.31$. $f_{{{\cal C}_0},C}$ peaks at $S=0.4$ akin to Fig. \ref{fig:Coop}.  $f_{{{\cal C}_0},C}$ rather than $f_{{{\cal C}_{(0,1]}},C}$ decides $f_{C}$ as seen in (b). ${\langle {\cal C} \rangle}_{{\cal H}_C}$ increases monotonically with $S$ in (c). Results are for  $f_{C_i}=0.5$, ${\cal N}=1024$, $\langle k\rangle=4$, $E_{\cal N} =1000$. The standard error is smaller than the size of the data points.} 
\label{fig:cc_all}
\end{figure}

We also study the clustering coefficient, ${\cal C}_i$, of node, $i$, at different values of $S$. We denote the total {\em number} of nodes in the network possessing ${\cal C}_i=0$ and $0<C_i\le 1$ by ${\cal N}_{{\cal C}_0}$  and ${\cal N}_{{\cal C}_{(0,1]}}$ respectively. The {\em fraction} of nodes in the network possessing ${\cal C}_i=0$ and $0< C_i\le 1$ is denoted by $f_{{\cal C}_0} ={\cal N}_{{\cal C}_0}/{\cal N}$ and $f_{{\cal C}_{(0,1]}} = {\cal N}_{{\cal C}_{(0,1]}}/{\cal N}$ respectively.  These numbers and fractions obviously include both cooperators and defectors. We now specifically denote the {\em number} of cooperators in the network possessing  ${\cal C}_i=0$ and $0< C_i\le 1$ by  ${\cal N}_{{\cal C}_0,C}$ and ${\cal N}_{{\cal C}_{(0,1],C}}$ respectively. The {\em fraction} of such nodes can then be respectively denoted by $f_{{\cal C}_0, C} ={\cal N}_{{\cal C}_0,C}/{\cal N}$ and $f_{{\cal C}_{[0,1)},C} = {\cal N}_{{\cal C}_{(0,1]},C}/{\cal N}$.

Fig. \ref{fig:cc_all}(a) exhibits the variation of $f_{{\cal C}_0,C}$ and $f_{{\cal C}_{(0,1]},C}$ versus $S$ at ${\cal T}=1.31$. We observe that the position of the peak for $f_{{\cal C}_0,C}$ mirrors that of $f_C$ as observed in Fig. \ref{fig:Coop} earlier.  Fig. \ref{fig:cc_all}(b) records the variation of $f_{{\cal C}_0}$ and $f_{{\cal C}_{(0,1]}}$ versus $S$. Clearly, $f_{{\cal C}_0}$  is far influential as compared to $f_{{\cal C}_{(0,1]}}$ in deciding $f_C$.  We have represented hubs acting as cooperators by ${{\cal H}_C}$. ${\langle {\cal C} \rangle}_{{\cal H}_C}$ denotes their average clustering coefficient. Fig. \ref{fig:cc_all}(c) demonstrates that ${\langle {\cal C} \rangle}_{{\cal H}_C}$ increases monotonically with $S$. 

The variation of ${\langle {\cal C} \rangle}_{{\cal H}_C}$ with respect to $S$ in Fig. \ref{fig:cc_all}(c)  is in remarkable contrast to the variation of ${\cal H}_C$ versus $S$ as observed in Fig. \ref{fig:hub}. As aforementioned, it has been reported earlier that the average clustering coefficient of a network is considered to work in favour of cooperation. However, we observe that the average clustering coefficient of hubs may not really promote cooperation. As $S$ increases, ${\langle {\cal C} \rangle}_{{\cal H}_C}$ increases monotonically, while the maintenance of cooperation progressively decreases. Indeed at $S=0.99$, ${\langle {\cal C} \rangle}_{{\cal H}_C}$ is at its highest yet the maintenance of cooperation is minimal. 

In order to gain a better understanding into the maintenance of cooperation, we take recourse to toy networks. In all toy networks considered herein;  ${\cal R}=1$, ${\cal T}=1.01$, ${\cal P}=0$, ${\cal S}=0$ \cite{santos2005scale}. Let $i$ and $j$ be two randomly chosen neighbors in the population. Let $A$ and $B$ denote the strategy of $i$ and $j$ respectively. This strategy can be either cooperation or defection. Let $k_i$ denote the degree of $i$, and, $k_j$ of $j$. If $k_{i_C}$ and $k_{i_D}$ be the number of $C$ and $D$ in the neighborhood of $i$, then $k_{i_C} + k_{i_D}=k_i$. Similarly, $k_{j_C} + k_{j_D}=k_j$. The accumulated payoff  of $i$ is
\begin{equation}
{\Pi}_i = \sum_{j} \pi_{ij} = {k_{i_C}}({\pi}_{C-A})+{k_{i_D}}({\pi}_{D-A})
\end{equation}
Obviously $A$ can be either $C$ or $D$. ${\pi}_{C-C} ={\cal R}$ (reward), ${\pi}_{C-D}={\cal T}$ (temptation), ${\pi}_{D-C}={\cal S}$ (sucker's payoff) and ${\pi}_{D-D}={\cal P}$ (punishment).The accumulated payoff of an arbitrarily chosen neighbor, $j$, of node, $i$, is
\begin{equation}
{\Pi}_j ={k_{j_C}}({\pi}_{C-B})+{k_{j_D}}({\pi}_{D-B})
\end{equation}
Individual, $i$, would upgrade to the strategy of $j$ with a probability $P(i \to j)$ as shown in Eqn. \ref{eq:P}. Similarly, $j$ can also imitate the strategy of $i$, with probability, ${P}_{j\rightarrow i}=\frac{\Pi_i-\Pi_j}{({\cal T}-{\cal S}) \times max(k_i, k_j)} \Theta_{\Pi_i \textgreater \Pi_j}$.

\begin{figure}[htbp]
\centering
\includegraphics[width=5cm,height=5.5cm]{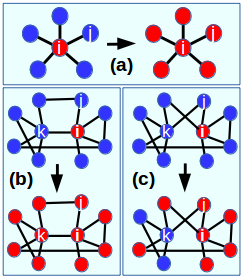}
\caption{Blue and red denote cooperators and defectors respectively. (a) Star graph with $({\cal N}, {\cal E})=(6,5)$. It can withstand the invasion of defection if the hub is not a defector. Graphs in (b) and (c) have $({\cal N}, {\cal E})=(10, 13)$ with $S_b \textgreater S_C$ and $r_b \textgreater r_C$. A direct link between two hubs, $i$ and $k$, makes this network vulnerable to defection. (b) Defection is likely to dominate if $i$ is a defector.  $i$ can turn $k$ into a defector. (c) Cooperation is possible as $k$ will never adopt defection.} 
\label{fig:toy}
\end{figure}

The star graph in Fig. \ref{fig:toy}(a) has one hub and five leaves. Suppose the hub, $i$, is a defector and the leaves are cooperators. Let $j$ be any arbitrarily chosen neighbor of $i$. Then, $A=D$, $B=C$, $k_{i_C}=5$, $k_{i_D}=0$, $k_{j_C}=0$, $k_{j_D}=1$. The accumulated payoff of $i$ and $j$ is $\Pi_i=5.05$, $\Pi_j=0$. Since ${\Pi_i >\Pi_j}$, $i$ will not imitate the strategy of its neighbor $j$. However, $j$ will imitate the strategy of $i$ with the probability $P(j\to i)=1$. Hubs play a significant role in the maintenance of cooperation. If the hub is a cooperator, it will acquire a higher payoff and gain an evolutionary advantage over its neighbors. However, maintenance of cooperation becomes fragile if the hub is a defector.

From Fig. \ref{fig:toy}(a) we observe that the strategy of the hub would dominate the outcome. Cooperation would be maintained in a star graph when the hub is itself a cooperator. However, the presence of many cooperator hubs in a network is not enough in itself for maintaining cooperation. There are two hubs in both the toy networks in Fig. \ref{fig:toy}(b) and \ref{fig:toy}(c).  Fig. \ref{fig:toy}(b) indicates that if hub $i$ is a defector, $k$ might adopt defection with a probability, $P(k\to i)$. From Eqn. \ref{eq:P}, we have $P(k\to i)=0.207$. Therefore, despite the presence of the cooperator hub, defection is likely to dominate the population. Evidently, a crucial role is played by the edge between $i$ and $k$. Due to this connection, a defector hub can easily affect its neighboring cooperator hub. On the other hand, in Fig. \ref{fig:toy} (c), we observe that if a defector hub and a cooperator hub are directly connected to each other, the defector hub would not be able to affect the cooperator hub.

In summary, a significant body of study in literature states that scale-free networks can facilitate cooperation. Herein, we 
examine  the prisoner's dilemma game on scale-free networks. We demonstrate that identical power-law degree distributions and indeed even an identical power-law degree sequence may exhibit remarkably different outcomes with regard to  cooperation.Our results indicate the maintenance of cooperation could be higher in SF networks within a narrow range of $S$ . 

We review the correlation of assortativity among cooperators and maintenance of cooperation. For this we 
borrow the notion of {\em ``cooperator graph"}, $\cal G_C$,  and {\em ``defector graph"}, $\cal G_D$ \cite{quorum}. 
We measure assortativity between cooperators, $r_C$, through the help of ${\cal G}_C$. We observe that the maintenance of cooperation does not always arise as a direct consequence of the assortativity between them. From the nature of variation of $f_C$ versus $r$, we also observe that cooperation does not bear a linear relationship with $r$.

We also study the average clustering coefficient of the network at different values of $S$. It has been reported that clustering directly influences the maintenance of cooperation in a network. However, we observe that for scale-free graphs with identical degree sequence, cooperation may not really depend on clustering.

In addition, we evaluate the role of  hubs towards the maintenance of cooperation. In a heterogeneous population, cooperator hubs play a crucial role in accumulating higher payoffs. $\forall {\cal G} \in {\cal G}_{SF}$, cooperation does not depend merely on the number of hubs, but rather on those hubs which are cooperators. However, whether the hubs become  cooperators or defectors would depend on the topology of the network. It appears that hubs are more likely to be directly connected to each other in graphs associated with higher values of $S$. If a hub becomes a defector, then other hubs are also likely to start adopting defection. Therefore, at higher values of $S$, cooperation becomes rather fragile due to the presence of direct edges between hubs. It would be beneficial to focus on clustering coefficient of cooperator hubs or even the clustering coefficient of individual nodes over the average clustering coefficient of the network. We have observed that an increase in clustering coefficient of the hubs is antagonistic to the maintenance of cooperation. Therefore, the presence of hubs is also not enough in itself to enhance the stability of cooperation. In summary, we can conclude that our existing understanding regarding cooperation on heterogeneous networks needs considerable revision. We scrutinise SF networks possessing an identical degree sequence and therefore an identical degree distribution. This leads us to observe that a power-law degree distribution may not be sufficient in itself for the maintenance of cooperation. {Further, the average clustering coefficient and assortativity may not have as large an influence over maintenance of cooperation as previously thought.}
\vspace{-0.35cm}

\bibliography{S-Metric_EPL}
\bibliographystyle{eplbib}
\vspace{-0.35cm}

\end{document}